\documentclass[
 reprint,
 superscriptaddress,
 amsmath,amssymb,
 aps,
 floatfix
]{revtex4-2}

\usepackage{todonotes}
\usepackage{float}  

\usepackage{graphicx}
\usepackage{dcolumn}
\usepackage{bm}
\usepackage{physics}
\usepackage{csquotes}
\usepackage{mathtools}
\usepackage{amsfonts}
\usepackage{amssymb}
\usepackage{amsmath}
\usepackage[caption=false,labelformat=empty]{subfig}
\usepackage{diagbox}
\usepackage{dsfont}
\usepackage{hyperref}
\usepackage{placeins}
\usepackage{xspace}
\usepackage{nicematrix}
\NiceMatrixOptions{cell-space-limits = 2pt}
\hypersetup{linktocpage,colorlinks,citecolor={blue},pdfdisplaydoctitle=true,pdfpagemode=UseOutlines,bookmarksnumbered=true}
\usepackage{verbatim}

\newcommand*{\dgr}{\ensuremath{^{\dagger}}}

\newcommand*{\id}{\ensuremath{\mathds{1}}}
\newcommand*{\tran}{\ensuremath{^T}}
\newcommand*{\inv}{\ensuremath{^{-1}}}

\newcommand{\site}[1]{\ensuremath{\mathbf{#1}}}

\begin{document}
\newcommand{\aqa}{$\langle aQa ^L\rangle $ Applied Quantum Algorithms, Universiteit Leiden}
\newcommand{\lorentz}{Instituut-Lorentz, Universiteit Leiden, Niels Bohrweg 2, 2333 CA Leiden, Netherlands}
\newcommand{\ulm}{Institute for Complex Quantum Systems, Ulm University, 89069 Ulm, Germany}
\newcommand{\iqst}{Center for Integrated Quantum Science and Technology (IQST), Ulm-Stuttgart, Germany}

\title{Algorithmic Aspects of Gauged Gaussian Fermionic PEPS: Gauge Fixing and Translation Invariance}

\author{Itay Gomelski}
\affiliation{School of Physics and Astronomy, Tel Aviv University, Tel Aviv 6997801, Israel}
\author{Jonathan Elyovich}
\affiliation{School of Physics and Astronomy, Tel Aviv University, Tel Aviv 6997801, Israel}
\author{Ariel Kelman}
\affiliation{Racah Institute of Physics, The Hebrew University of Jerusalem, Givat Ram, Jerusalem 91904, Israel}
\author{Erez Zohar}
\affiliation{School of Physics and Astronomy, Tel Aviv University, Tel Aviv 6997801, Israel}
\author{Patrick Emonts}
\affiliation{\ulm}
\affiliation{\iqst}
\affiliation{\aqa}
\affiliation{\lorentz}

\begin{abstract}
Lattice gauge theories (LGTs) provide a powerful framework for studying non-perturbative phenomena in gauge theories. However, conventional approaches such as Monte Carlo (MC) simulations in imaginary time are limited, as they do not allow real time evolution and suffer from a sign problem in many important cases. Using Gauged Gaussian fermionic projected entangled pair states (GGFPEPS) as a variational ground state ansatz offers an alternative for studying LGTs through a sign-problem-free variational MC. As this method is extended to larger and more complex systems, understanding its numerical behavior becomes essential. While conventional action based MC has been extensively studied, the performance and characteristics of non-action-based MC within the GGFPEPS framework are far less explored. In this work, we investigate these algorithmic aspects,
identifying an optimal update size for GGFPEPS-based MC simulations for $\mathbb{Z}_2$ in $2+1$ dimensions. 
We show that gauge fixing generally slows convergence, and demonstrate that not exploiting the translation-invariance can, in some cases, improve the computational time scaling of error convergence. 
We expect that these improvements will allow advancing the simulation to larger and more complex systems.
\end{abstract}

\maketitle
\section{Introduction}
Gauge theories are fundamental in both particle physics and condensed matter physics. 
In particle physics they describe the fundamental forces of nature in the standard model~\cite{peskin_introduction_1995}. 
In condensed matter physics, they often appear as effective low-energy models for many-body phenomena such as quantum spin liquids~\cite{fradkin_field_2013,savary_quantum_2016-1}. 
In addition, quantum chromodynamics (QCD) is a gauge theory describing the theory of strong nuclear interactions. 
Due to asymptotic freedom, the coupling in QCD becomes small at high energies, allowing the use of perturbation theory~\cite{gross_asymptotically_1973}. 
In contrast, the low-energy regime is dominated by strongly non-perturbative phenomena and confining forces, so perturbation theory breaks down and alternative methods must be employed.
Nevertheless, this regime gives rise to some of the most fascinating aspects of QCD, such as quark confinement and hadronization~\cite{PhysRevD.10.2445quarkconfinement,GREENSITE20031confinementinlgt}.

One approach to access this non-perturbative regime is to discretize space or spacetime in the framework of lattice gauge theories~\cite{wilson_confinement_1974,kogut_hamiltonian_1975,kogut_introduction_1979}. Monte Carlo (MC) simulations in imaginary time can then be performed~\cite{creutz_monte_1980, creutz_monte_1983}.
Such simulations have been very successful in computing many static quantities in gauge theories~\cite{creutz_monte_1983}. 
However, this method faces two main limitations. 
First, working in a Euclidean spacetime prevents the study of real-time dynamics, as the time evolution operator becomes a decaying exponential. 
Second, in many physically relevant cases involving finite fermionic density, the sign problem arises --- the probabilistic interpretation required for MC simulations breaks down because the probability weight becomes complex or negative.

Working in the Hamiltonian formalism with discretized space and continuous time~\cite{kogut_hamiltonian_1975} addresses the issue of real-time evolution. However, the corresponding Hilbert space can be infinite-dimensional for continuous gauge groups (and still grows exponentially with system size even for finite groups), making exact solutions impossible. Variational methods offer a viable alternative, with the main challenge being the construction of a suitable ansatz.

Tensor network states, particularly, matrix product states (MPS) and their higher-dimensional generalizations, such as projected entangled pair states (PEPS), are many-body states built from locally entangled degrees of freedom. They can efficiently approximate ground states and thermal states of local gapped Hamiltonians~\cite{cirac_renormalization_2009,ORUS2014117,cirac_PEPS_general}. 
Each lattice site hosts a physical degree of freedom, while the links carry virtual ones. 
The virtual degrees of freedom are entangled between neighboring sites, and contracting them over the entire network yields a physical state. 
Tensor network states can satisfy by construction entanglement entropy area law and allow one to naturally encode symmetries. 
These properties make them an appropriate ansatz for the ground state of local gapped Hamiltonians
\cite{hastings_area_2007,white_density_1992,fannes_finitely_1992,Eisert_area_law}.

In one spatial dimension, MPS combined with density matrix renormalization group (DMRG)~\cite{white_density_1992,schollwock_density-matrix_2011} techniques have been successfully applied to lattice gauge theories, reproducing known results and providing new insights into systems that suffer from sign problems in imaginary time MC simulations~\cite{banuls_mass_2013,buyens_matrix_2014,rico_tensor_2014,kuhn_non-abelian_2015,banuls_thermal_2015,pichler_real-time_2016,buyens_hamiltonian_2016,dalmonte_lattice_2016,banuls_simulating_2020,banuls_review_2020}. 
However, in higher dimensions, contracting the state scales very poorly with system size~\cite{schuch_computational_2007}. 
To overcome these limitations, studies have turned to alternative variational ans\"atze, ranging from advanced tensor network structures, such as tree tensor networks and infinite PEPS~\cite{tagliacozzo_entanglement_2011,tagliacozzo_tensor_2014,crone_detecting_2020,robaina_simulating_2021,felser_two-dimensional_2020,magnifico_lattice_2021,montangero_loop-free_2022,cataldi_21d_2023} to neural network wave functions~\cite{priggs_2025_neural_LGT}.

Recently, a new type of tensor network state has been proposed that combines the advantages of tensor network states, while providing an alternative for contracting them. 
These states, known as gauged Gaussian fermionic PEPS (GGFPEPS)~\cite{MC_zohar_2018,emonts_2020,emonts2023,kelman2024gauged,kelman2024projectedentangledpairstates}, serve as variational MC ansatz for LGTs ground states in more than one spatial dimension. 
They are designed to respect global symmetries of the Hamiltonian as well as gauge invariance through a gauging transformation. 
GGFPEPS exhibit advantages of tensor network states, such as the entanglement entropy area law, while enabling efficient computation of norms and expectation values through the covariance matrix formalism~\cite{Bravyi-gaussianformalism} (thanks to their Gaussian nature) and sign problem free MC simulations.

So far, the GGFPEPS have been successfully applied to numerically compute the ground state of a pure gauge $\mathbb{Z}_3$~\cite{emonts_2020} LGT and a $\mathbb{Z}_2$ LGT both with~\cite{kelman2024projectedentangledpairstates} and without~\cite{emonts2023} dynamical fermionic matter. 
The next steps in this line of research involve extending the method to higher spatial dimensions or more complex gauge groups, such as compact Lie and non-Abelian groups. 
However, as systems become larger and more complicated, the numerical and algorithmic details of the simulations play an increasingly critical role in obtaining meaningful results.

In this paper, we study several aspects of MC simulations within the GGFPEPS framework. 
Although conventional action-based MC methods are well understood, the behavior and performance of non-action-based MC within the GGFPEPS framework remain much less explored. 
In particular, we investigate how the error convergence rate of the MC simulation changes as a function of different simulation parameters, as well as suggest new approaches to improve the MC simulation. 
We identify an optimal update size for MC simulations with GGFPEPS. 
We suggest exploiting the ansatz's gauge invariance by fixing the gauge and reducing the number of degrees of freedom when computing expectation values of physical observables. 
When contracting GGFPEPS exactly, this reduces the complexity of evaluating physical observables by roughly a square-root factor.
However, we observe that gauge fixing generally slows MC convergence, and we provide an explanation for this behavior.
Moreover, we find that not relying on the translation invariance assumption can, in some cases, improve error convergence in MC simulations over time.

The rest of this work is structured as follows:
sections \ref{background_and_ansatz}, \ref{sec:algorithm} and \ref{section:cov_mat} present the LGT Hamiltonian and the GGFPEPS method, including the construction of the ansatz and its use in Monte Carlo simulations within the Gaussian formalism --- these sections provide an overview of the work in Ref.~\cite{kelman2024gauged}. 
In Section~\ref{sec:MC_errors} we briefly discuss error computations in the MC simulations and different approaches to estimate them. 
Later, the dependence of the MC error convergence on the update-step size is examined in Section~\ref{sec:update_size}. In Section~\ref{sec:gauge_fixing} we show that it is possible to fix the gauge in our framework and examine its effect on the MC convergence rate. 
Section~\ref{sec:TI}, studies the effect of not exploiting translation invariance on the MC the error convergence. 
Finally, Section~\ref{sec:conclusions} summarizes the results and conclusions.

Throughout this entire work, we assume summation over repeated indices, unless they represent an irreducible group representation. 
The ansatz construction is presented for a general gauge group to maintain the generality of Section~\ref{sec:gauge_fixing}, even though all numerical results correspond to a $\mathbb{Z}_2$ gauge theory with dynamical fermions (for details on this system see Ref.~\cite{kelman2024projectedentangledpairstates}).

\section{Physical Background and ansatz} \label{background_and_ansatz}
This section provides an overview of the Hamiltonian formulation of lattice gauge theory for a general gauge group. We also present the construction of an ansatz for the ground state of the theory in the form of a GGFPEPS.

\begin{figure}
    \centering
	\includegraphics[width=0.3\textwidth]{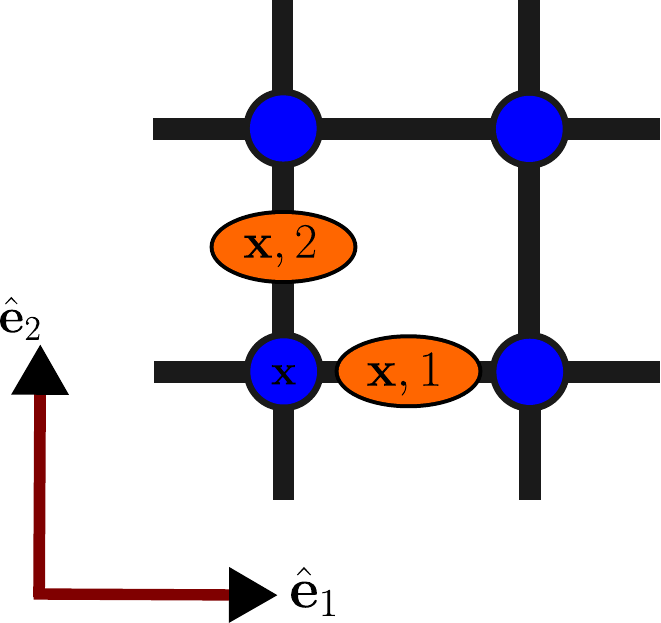}
	\caption{Illustration of the lattice and the labeling convention for sites and links.}
    \label{fig:lattice_scheme}
\end{figure}

The Hilbert space of a lattice gauge theory consists of two different types of degrees of freedom --- fermionic matter residing on the lattice's sites, and gauge fields residing on the links. 
Here, we consider a square lattice in $2$ spatial dimensions with periodic boundary conditions. 
Lattice sites are denoted by $\site{x}=\left(x_1,x_2\right)$ and links by $\ell=\left(\site{x},i\right)$, where $i=1,2$ is the lattice dimension (see Fig.~\ref{fig:lattice_scheme}). 

The local Hilbert space of the gauge fields is spanned by the group element basis (magnetic basis) $\left\{\ket{g}\right\}_{g \in G}$, labeled by elements of the gauge group $G$. 
The group $G$ is either a finite group or a compact Lie group. 
On each link, we introduce right and left group operations represented by the unitary operators \(\Theta_g\) and \(\widetilde{\Theta}_g\), respectively, such that
\begin{equation} \label{eq:group_op}
    \Theta_g \ket{h} = \ket{h g^{-1}} \qquad \text{and} \qquad \widetilde{\Theta}_g \ket{h} = \ket{g^{-1}h}.
\end{equation}

We define the group element operator as
\begin{equation}
    U^j_{mn}=\int dg \, D^j_{mn}(g) \ket{g}\bra{g}.
\end{equation}
It transforms under group operations as 
\begin{align}
    \Theta_g U^j_{mn} \Theta^{\dagger}_g &= U^j_{mn'} D^j_{n'n}(g), \\
    \widetilde{\Theta}_g U^j_{mn} \widetilde{\Theta}^{\dagger}_g &= D^j_{mm'}(g) U^j_{m'n}.
\end{align}
where $D^j (g)$ is a unitary $j$ representation of $g$ (Wigner matrix).
Alternatively, the local Hilbert spaces can be spanned by the representation basis (electric basis) \(\ket{jmn}\), where \(j\) represents an irreducible representation (irrep), and \(m\) and \(n\) are multiplet indices corresponding to the eigenvalues of the maximal set of mutually commuting operators for the left and right transformations, respectively. 

On each site $\site{x}$ we introduce the fermionic creation operators $\psi^\dagger_{m}(\site{x})$, where $m$ is a color index which determines how the fermionic modes interact with the gauge field. 
Additional properties such as flavor can be straightforwardly added, as done in~\cite{kelman2024gauged}. 
To address the lattice fermions doubling problem, we will consider here a staggered fermionic picture~\cite{susskind_fermions}, splitting the sites (vertices) into two sublattices --- even and odd, each hosting different spinor component. 
Alternative conventions can also simplify this problem, such as placing fermions with multiple spin components on each site. 
Under a gauge operation $g\in G$ operations, the matter transforms under a unitary operator $\theta_g$,
\begin{equation}
\theta_g\left(\site{x}\right)
\psi^\dagger_{m}(\site{x})
\theta_g^{\dagger}\left(\site{x}\right)
= \psi^\dagger_{n}(\site{x})D^j_{nm}\left(g\right),
\end{equation}
where here and throughout the rest of the work, the irrep index $j$ of the matter fields is omitted.

We now have all the ingredients to construct the LGT Hamiltonian, which is composed of several parts --- the fermionic part, $H_F$ and gauge field dynamics which are given by electric energy, $H_\text{E}$ and magnetic energy, $H_\text{B}$, terms,
\begin{equation}
    H = H_F + H_{\text{E}} + H_{\text{B}}.
\end{equation}
The fermionic Hamiltonian $H_F$ is of the form,
\begin{equation} \begin{aligned}
	H&_F = M\underset{\site{x}}{\sum}\left(-1\right)^{\site{x}}\psi_{m}^{\dagger}\left(\site{x}\right)\psi_{m}\left(\site{x}\right)  \\&+
		\frac{i}{2a}\Bigg(
		 \sum_{\site{x}} \bigg[ \psi_{m}^{\dagger}\left(\site{x}\right)
   U_{mn}\left(\site{x},1\right)
   \psi_{n}\left(\site{x}+\hat{\mathbf{e}}_1\right)
		\\ &+
  i\left(-1\right)^{\site{x}} \psi_{m}^{\dagger}\left(\site{x}\right)
        U_{mn}\left(\site{x},2\right)
    \psi_{n}\left(\site{x}+\hat{\mathbf{e}}_2\right) \bigg]
		-\text{h.c.}
		\Bigg).
\end{aligned} \end{equation}
The gauge field dynamics are described by the Kogut–Susskind Hamiltonian~\cite{PhysRevD.11.395_wilson_lattice}, which consists of magnetic, $H_\text{B}$, and electric, $H_\text{E}$, energy terms, 
\begin{equation}
    H_{\text{KS}} = H_{\text{E}} + H_{\text{B}}.
\end{equation}
Here, $H_{\text{E}}$ is the electric energy operator, defined as:
\begin{equation}
    H_{\text{E}} = g_{\text{E}} \sum_{\left(\site{x},i\right),j} f_j P_j\left(\site{x},i\right),
\end{equation}
where $g_{\text{E}}$ and $f_j$ are real parameters, $P_j(\site{x},i) = \ket{jmn}  \bra{jmn}$  is a projector onto the $j$ irrep on a link $(\site{x},i)$. 
The magnetic term \(H_{\text{B}}\) is given by:
\begin{equation} \begin{aligned}
    H_{\text{B}} &= g_{\text{B}}\sum_{\site{x}}
    \bigg( \text{Tr} \Big[ U\left(\site{x},1\right)U\left(\site{x}+\hat{\mathbf{e}}_1,2\right)
    \\ &\quad \times
    U^{\dagger}\left(\site{x}+\hat{\mathbf{e}}_2,1\right)U^{\dagger}\left(\site{x},2\right)\Big] + \text{h.c.} \bigg).
\end{aligned} \end{equation}
where the irrep index is omitted, and is the same as for the matter.

Additionally, a chemical potential term can be included and accordingly the ansatz can be straightforwardly generalized as done in Ref.~\cite{kelman2024gauged}.

The Hamiltonian has several symmetries. 
Most importantly, it is invariant under local gauge transformations $\hat{\Theta}_g$ that include a lattice site and its surrounding links,
\begin{equation}\label{eq:thetahat}
    \hat{\Theta}_g(\site{x})=\prod_{i=1,2} \Tilde \Theta_g(\site{x},i) \Theta_g^\dagger(\site{x}-\hat{\mathbf{e}}_i,i)\theta^{\dagger}_g(\site{x}),
\end{equation}
i.e. $\left[H,\hat{\Theta}_g(\site{x})\right]=0$
for every lattice site $\site{x}$ and $g \in G$. 
This symmetry splits the Hilbert space into sectors classified by $\hat{\Theta}_g(\site{x})$ eigenvalues. 
We focus on the sector with no static charges corresponding to states $\ket{\psi}$ satisfying
\begin{equation} \label{eq:gauge_inv_state}
    \hat{\Theta}_g(\site{x}) \ket{\psi}=\ket{\psi}.
\end{equation}
Specifically, the states we are interested in are strictly gauge invariant and satisfy Eq.~\eqref{eq:gauge_inv_state}.

The Hamiltonian also has a global $U(1)$ symmetry generated by the total fermionic number $N_0=\sum_\site{x}\psi_m\dgr(\site{x})\psi_m(\site{x})$. 
Moreover, it exhibits invariance under two site translations, which is characteristic of staggered fermions and has a $\frac{\pi}{2}$ rotation symmetry. 

We would like to construct a PEPS which has the Hamiltonian's global symmetries and is invariant under gauge transformations, i.e. satisfies Eq.~\eqref{eq:gauge_inv_state}.

As mentioned previously, PEPS include both virtual and physical degrees of freedom, with the virtual ones contracted to mediate interactions between neighboring sites.
Since these virtual degrees of freedom will be coupled to physical fermions, we define them as fermionic modes. 
For each site $\site{x}$ we associate several copies of virtual modes on each of the four legs of the site, denoted by $a^{\dagger}_{i \mu m}\left(\site{x}\right)$. 
Here, $m$ denotes the color index, $i \in \{1, 2, 3, 4\}$ corresponds to the different legs $\hat{\mathbf{e}}_1$, $\hat{\mathbf{e}}_2$, $-\hat{\mathbf{e}}_1$, $-\hat{\mathbf{e}}_2$, respectively, and $\mu$ labels different copies, allowing for multiple virtual modes per leg. 
Varying the number of copies controls the expressibility of the ansatz (modifies the bond dimention of the PEPS).
Out of these modes we construct the Gaussian operator
\begin{equation}
    A\left(\site{x}\right)=\exp \left(\mathcal{T}_{\alpha \beta} \Psi^\dagger_\alpha(\site{x}) \Psi^\dagger_\beta(\site{x})\right),
\end{equation}
 where $\Psi^\dagger_\alpha (\site{x}) 
    \in \{ \psi^\dagger_m (\site{x}) \} \cup \{ a^{\dagger}_{i \mu m}(\site{x}) \}$ (with $\alpha$ containing all different indices), and $\mathcal{T}(\site{x})$ is a complex matrix coupling different fermionic modes. 

Next, we define the projection operator, which is a Gaussian operator composed of only virtual modes, 
\begin{equation}
    w\left(\site{x},k\right)=\exp \left(X^{(k)}_{ij} 
\underset{\mu}{\sum} W^{(i, \mu)}
a^{\dagger}_{i\mu m}\left(\site{x}\right)
a^{\dagger}_{j\mu m}\left(\site{x}+\hat{\mathbf{e}}_k\right)\right),
\end{equation}
where there is no summation over $k$, and
$X^{(1)}_{ij} =\delta_{i,1}\delta_{j,3}, \, X^{(2)}_{ij} =\delta_{i,2}\delta_{j,4}$. 
As we will see, the Gaussian nature of $A$ and $w$ enables efficient computation of physical observables, since all information about Gaussian states is  encoded in their covariance matrices.

Thus we can construct the following PEPS,
\begin{equation}
\ket{\psi_0} = 
\bra{\Omega_\text{v}} \prod_{(\site{x}, i)}
w^{\dagger}\left(\site{x}, i \right)
\underset{\boldsymbol{x}}{\prod}A\left(\site{x}\right)\ket{\Omega_{\text{p}}}\ket{\Omega_{\text{v}}},
\label{PEPSdef}
\end{equation}
where $\Omega_\text{v}$ and $\Omega_\text{p}$ are the virtual and physical Fock vacua.
One can now impose the global symmetries of the Hamiltonian --- $\frac{\pi}{2}$ rotations, $U\left(1\right)$, translation invariance and global $G$. 
This places constraints on the coupling matrices in the $A$ and $w$ operators. 
For details on the explicit parametrization and the implementation of global symmetries see Ref.~\cite{kelman2024gauged} for a general gauge group, or  Refs.~\cite{emonts2023,kelman2024projectedentangledpairstates} for the $\mathbb{Z}_2$ case.

This leads to the final stage of the ansatz construction which is gauging, i.e., adding gauge field degrees of freedom and promoting the global $G$ symmetry to a local gauge invariance. 
This is done using a controlled operation $U_G$ entangling the virtual fermions with the gauge fields lifting the global $G$ symmetry into a local gauge symmetry. 
Thus, the state takes its final form,
\begin{equation}
\ket{\Psi}=\bra{\Omega_{\text{v}}} \underset{\site{x},i}{\prod}w^{\dagger}\left(\site{x},i\right)
\mathcal{U}_G\left(\site{x},i\right)
\underset{\site{x}}{\prod}A\left(\site{x}\right)\ket{\Omega_{\text{p}}}\ket{\Omega_{\text{v}}} \ket{\site{0}},
\label{GPEPSdef}
\end{equation}
where $\ket{\site{0}}\equiv\underset{\site{x},i=1,2}{\bigotimes}\ket{000}$, and $\ket{000}$ denotes the $j=0$ singlet state in the representation basis. 

We expand the state in the group element configuration basis,
\begin{equation}\label{eq:psi_group_exp}
    \ket{\Psi} = \int \mathcal{DG} \ket{\mathcal{G}}
    \ket{\psi\left(\mathcal{G}\right)},
\end{equation}
where,
\begin{equation}
\ket{\psi\left(\mathcal{G}\right)}=\bra{\Omega_{\text{v}}}\underset{\site{x},i}{\prod}w^{\dagger}\left(\site{x},i\right)\underset{\site{x},i}{\prod}\mathcal{U}_{g}\left(\site{x},i\right)\underset{\site{x}}{\prod}A\left(\site{x}\right)\ket{\Omega_{\text{p}}}\ket{\Omega_{\text{v}}}
\label{eq:psi(g)}
\end{equation}
is a matter state coupled to a gauge field configuration $\mathcal{G}$. 
The measure $\mathcal{DG}$ is defined as,
\begin{equation}
    \mathcal{DG}=\underset{\site{x},i}{\prod}dg\left(\boldsymbol{x},i\right),
\end{equation}
where $dg$ is the Haar measure. 
For compact Lie groups, $dg$ corresponds to an integration measure over the group manifold, whereas for discrete groups the integral reduces to a sum over the group elements.
The operator $\mathcal{U}_{g}\left(\site{x},i\right)$ is the fermionic part of the gauging transformation, satisfying
\begin{equation}
    \int dg \ket{g}\bra{g}_{\site{x},i}\otimes \mathcal{U}_{g}\left(\site{x},i\right)  =U_G(\site{x},i).
\end{equation}

Note that for a fixed gauge field configuration $\mathcal{G}$, the coupled matter state $\ket{\psi(\mathcal{G})}$ is Gaussian.
This expansion will be useful in the next section when we discuss calculation of expectation values of physical observables.

\section{The algorithm} \label{sec:algorithm}
After constructing the ansatz state $\ket{\Psi}$, the next step is to perform a variational Monte Carlo ground state search by minimizing the expectation value of the Hamiltonian with respect to the parameters in the $\mathcal{T}$ matrix. 
To compute the Hamiltonian’s expectation value, we must evaluate the physical gauge-invariant observables on which it depends.
In order to calculate the expectation values we will use $\ket{\Psi}$'s group configuration expansion in Eq.~\eqref{eq:psi_group_exp}.
Consider some gauge invariant operator $\mathcal{O}$, its expectation value with respect to the ansatz is
\begin{equation} \label{eq:general_exp_val}
\left\langle \mathcal{O} \right\rangle =
\int \mathcal{DG} \ F_{\mathcal{O}}\left(\mathcal{G}\right) p\left(\mathcal{G}\right),
\end{equation}
where $F_\mathcal{O}(\mathcal{G})$ is an observable dependent function of the gauge field configuration and
\begin{equation}\label{eq:prob_dist}
    p\left(\mathcal{G}\right) = \frac{\braket{\psi\left(\mathcal{G}\right)}}{\int \mathcal{D}\mathcal{G}' \braket{\psi\left(\mathcal{G}'\right)}} \equiv \frac{p_{0}\left(\mathcal{G}\right)}{\mathcal{Z}}.
\end{equation}
Note that $p(\mathcal{G})$ is always a valid probability distribution since it is real, positive and normalized. Thus this expectation value can be calculated using sign-problem-free Monte Carlo sampling.
Since $\ket{\psi\left(\mathcal{G}\right)}$ is Gaussian, its norm $p_0\left(\mathcal{G}\right)$ can be computed efficiently in a covariance matrix formalism. Furthermore, for observables of interest to us (gauge invariant that appear in the Hamiltonian), $F_\mathcal{O}(\mathcal{G})$ consists of expectation values of physical observables with respect to $\ket{\psi\left(\mathcal{G}\right)}$~\cite{kelman2024gauged}, and can therefore be computed efficiently using covariance matrix formalism~\cite{Bravyi-gaussianformalism}. 

Thus, for the evaluation of expectation values we use Markov chain Monte Carlo (MCMC)~\cite{metropolis_equation_1953}. 
We construct a Markov chain of gauge configurations with transition probabilities given by \eqref{eq:prob_dist}. We start with a warmup phase in which we set a random gauge configuration $\mathcal{G}$ and compute its weight $p_0(\mathcal{G})$ using covariance matrix formalism. 
We then randomly sample a new gauge configuration $\mathcal{G}'$ and compute its weight $p_0(\mathcal{G}')$.
This configuration is accepted with probability,
$\frac{p(\mathcal{G}')}{p(\mathcal{G})} = \frac{p_0(\mathcal{G}')}{p_0(\mathcal{G})} 
$ (in case this ratio is larger than $1$, it is accepted), otherwise, the configuration is rejected. 
This process is repeated for a predetermined number of steps, allowing the Markov chain to converge to the desired probability distribution. 
Once convergence is reached, we start with the measurement phase in which in addition to sampling and accepting/rejecting the proposed gauge configurations, we evaluate the value of $F_\mathcal{O}(\mathcal{G})$ for all observables and store it (in case we reject a proposed configuration, we store again the $F_\mathcal{O}$ value for the previous configuration). 
At the end of this phase, we compute the means of all the physical observables functions $F_\mathcal{O}$.

For GGFPEPS, the computation time of the warmup phase tends to be much shorter than for the measurement phase. 
Therefore, the MC analysis shown in the following sections focuses on the measurement phase after warmup.

For the variational search of the ground state, we compute the expectation value of the Hamiltonian and its gradient with respect to the ansatz parameters using an exact analytical expression, expressible in terms of derivatives of the relevant covariance matrices, as shown for the electric energy in~\cite{emonts2023}.
We then update these parameters to minimize the gradient, using gradient descent or a similar optimization algorithm. 
By iterating this process, we minimize the Hamiltonian's expectation value and reach an approximation of the ground state.

\section{Covariance matrix formalism} \label{section:cov_mat}
Here, we will go over details of the final ingredient of the algorithm, the covariance matrix formalism and its use in computing the weights. 
The following brief overview is intended to complete the details required for understanding the MC local updating scheme presented in section \ref{sec:update_size}. 
For more details we refer the reader to~\cite{Bravyi-gaussianformalism, MC_zohar_2018}

Let us consider $N$ fermionic modes $\{a_i \}_{i=1} ^N$. 
Our of these, one can construct $2N$ Majorana modes, $\gamma_i^{\left(1\right)}=a_i+a_i^\dagger$,  $\gamma_i^{\left(2\right)}=i(a_i-a_i^\dagger)$ which satisfy the anticommutation relation of the Clifford algebra, $\left\{\gamma_\alpha^{\left(1\right)},\gamma_\beta^{\left(2\right)}\right\}=2\delta_{\alpha \beta}$ .
The covariance matrix for a pure state $\ket{\phi}$ is defined as, 
\begin{equation}
     \Gamma_{i j} = \frac{i}{2}\frac{\bra{\phi}\left[\gamma_i,\gamma_j\right]\ket{\phi}}{\braket{\phi}}.
\end{equation}
For Gaussian $\ket{\phi}$, all information of the state is encoded in this matrix.

We denote the covariance matrix of the gauged projector $\ket{B(\mathcal{G})}=\underset{\site{x},k}{\prod}\mathcal{U}_g^\dagger\left(\site{x},i\right)w\left(\site{x},i\right)
\ket{\Omega_{\text{v}}}$, its density matrix by $\rho^B\left(\mathcal{G}\right)=\ket{B\left(\mathcal{G}\right)}\bra{B\left(\mathcal{G}\right)}$, and its covariance matrix by $\Gamma_\text{in}\left(\mathcal{G}\right)$. Furthermore, we denote the state $\ket{A}=\underset{\site{x}}{\prod}A\left(\site{x}\right)\ket{\Omega_{\text{p}}}\ket{\Omega_{\text{v}}}$, its density matrix by $\rho^A = \ket{A}\bra{A}$, and its covariance matrix by $M$.
Following Eq.~(\ref{eq:psi(g)}), the weights can now be presented as an overlap of two Gaussian states,
\begin{equation}
p_0\left(\mathcal{G}\right)=\braket{\psi\left(\mathcal{G}\right)}=\text{Tr}\left[\rho^B\left(\mathcal{G}\right)\otimes \id_\text{physical}\rho^A\right],
\end{equation}
where the product with the identity operator of the physical modes $\id_\text{physical}$ embeds $\rho^B$ into the Hilbert space of the physical modes.
This overlap can be computed using covariance matrices~\cite{Bravyi-gaussianformalism,kelman2024gauged,MC_zohar_2018},
\begin{equation} \label{eq:norm_cov}
    \braket{\psi\left(\mathcal{G}\right)} \propto
    \sqrt{\det \bigg(\frac{1-\Gamma_{\text{in}}\left(\mathcal{G}\right)D}{2} \bigg)}
\end{equation}
where $D$ denotes the block of $M$ which couples virtual fermions to themselves. 
Thus, the weights can be computed efficiently within the covariance matrix formalism, utilizing the Gaussian nature of $\ket{\psi(\mathcal{G})}$. 
Using similar tools, one can also evaluate the observable function $F_\mathcal{O}\left(\mathcal{G}\right)$.

\section{Error estimation in Monte Carlo simulations} \label{sec:MC_errors}
Our analysis focuses on the Monte Carlo error convergence under different simulation settings.
To this end, we first outline how the errors are computed.

Consider a set of $N$ MC measurements for an observable $\mathcal{O}$ (after the warmup phase), $\left\{F_\mathcal{O}\left(\mathcal{G}_i\right)\right\}_{i=1}^N$. 
In case of independent and identically distributed (IID) samples the error on the mean (EOM) is estimated by
\begin{equation} \label{eq:IID_EOM}
  \text{EOM}_{\text{IID}}\left(F_\mathcal{O}\right)=\sqrt{\frac{\text{Var}\left(F_\mathcal{O}\right)}{N}},
\end{equation}
where $\text{Var}\left(F_\mathcal{O}\right) = \left\langle F_\mathcal{O}^2 \right\rangle - \left\langle F_\mathcal{O} \right\rangle^2$ is the variance of the $F_\mathcal{O}$ samples and $\left\langle F_\mathcal{O}\right\rangle=\frac{1}{N}\sum_{i=1}^N F_\mathcal{O}\left(\mathcal{G}_i\right)$ is the sample mean~\cite{newman1999monte}.

However, in a Markov chain, the samples are correlated and thus the error cannot be computed in such a straightforward manner. The EOM of correlated samples can be estimated by~\cite{janke2002statistical},
\begin{equation}\label{eq:EOM_autocorr}
    \left(\text{EOM}\left(F_\mathcal{O}\right)\right)^2=\frac{\text{Var}\left(F_\mathcal{O}\right)}{N}\left(1+2\sum^N_{t=1}A_{F_\mathcal{O}}\left(t\right)\left(1-\frac{t}{N}\right)\right),
\end{equation}
where $A_{F_\mathcal{O}}\left(t\right)$ is the autocorrelation which quantifies the correlation between samples samples separated by $t$ steps,
\begin{equation}
    A_{F_\mathcal{O}}\left(t\right)=\frac{\left\langle F_\mathcal{O}\left(\mathcal{G}_i\right)F_\mathcal{O}\left(\mathcal{G}_{i+t}\right)\right\rangle}{\left\langle F_\mathcal{O}\left(\mathcal{G}\right)^2 \right\rangle- \left\langle F_\mathcal{O}\left(\mathcal{G}\right)\right\rangle^2}.
\end{equation}
The averages are taken over all available pairs $\left(i,i+t\right)$ in the series, and the denominator ensures the normalization $A_{F_\mathcal{O}}(0)=1$.

The autocorrelation function typically decays exponentially as can also be seen in our simulation, in Figs.~\ref{fig:binning_update_size} and~\ref{fig:gf_combined}. 
In our simulation, after a significant decay to about $10^{-3}$, the autocorrelation begins to fluctuate between $10^{-2}$ and $10^{-3}$.

This allows us to define a decay time $\tau$ (in terms of MC steps), as the number of steps required for the autocorrelation to decrease by a significant factor (we use $10^{-2}$). 
Thus, we can rebin the data by grouping each $\tau$ consecutive samples and averaging over them, resulting in $n_B = \left\lfloor \frac{N}{\tau} \right\rfloor$ bins~\cite{Sandvik_spin_autocorr},
\begin{equation}
    F_{\mathcal{O},\text{rebinned}}^i=\sum_{j=\left(i-1\right)n_B+1}^{i\cdot n_B} F_\mathcal{O}\left(\mathcal{G}_j\right).
\end{equation}
Consequently, we effectively obtain $n_B$ samples of less correlated data $\left\{F_{\mathcal{O},\text{rebinned}}^i\right\}_{i=1} ^{n_B}$, and following Eq.~\eqref{eq:IID_EOM} the error on the mean can be computed as follows,
\begin{equation}\label{eq:EOM}
  \text{EOM}\left(F_\mathcal{O}\right)=\sqrt{\frac{\text{Var}\left(F_{\mathcal{O},\text{rebinned}}^i\right)}{n_B}}.
\end{equation}
Note that unlike variance and EOM, the mean remains invariant after rebinning.
This allows for a much more practical approach to computing the EOM than Eq.~\eqref{eq:EOM_autocorr}. Throughout this paper, error on the mean (EOM) refers to Eq.~\eqref{eq:EOM}.

This implies that the autocorrelation is closely related to the error --- the faster it decays, the smaller $\tau$ becomes and the more bins can be formed. 
Moreover, the variance itself depends on the chosen rebinning.

\section{Update Size} \label{sec:update_size}
In MC evaluations there are several approaches for updating the gauge field configurations --- a local update which modifies a fixed, small number of links, and a global update which involves changing a large amount of links (usually a constant fraction of the system size) \cite{wolf_cluster,swendsen_wang_cluster}. 
Global updates typically achieve faster error convergence (in terms of step number) and have better ergodicity compared to local updates. 
However, not only are global updates generally more computationally expensive per step, but constructing an unbiased scheme is often highly nontrivial.

Naively speaking, for any type of MC update, reevaluating the weight $p_0(\mathcal{G})$ and the observable's function $F_\mathcal{O}(\mathcal{G})$ for the updated gauge field configuration, requires recalculating the determinant and the inverse of matrices whose size matches that of the covariance matrices for the whole system (see Eq.~\eqref{eq:norm_cov}). 
For a $d$ dimensional lattice with linear length $L$, their size scales like $L^d$, which means that a naive update, regardless of the amount of links it involves, could be extremely costly numerically. 
However, we use a local update scheme which enables a speed-up, using the Woodbury matrix identity~\cite{higham_accuracy_2002} and the matrix determinant lemma~\cite{harville_matrix_1997}.

The matrix determinant lemma is used to compute the determinant of a matrix $M$ after a local update $C$,
\begin{equation}
  \det(M+UCV\tran) = \det(C\inv+V\tran M\inv U) \det(C) \det(M),
  \label{eq:det_lemma}
\end{equation}
where $U$ and $V$ consist only of $0$ or identity blocks, and are used to place the update $C$ to match the dimensions of $M$. 
In our case $M=\left(D\inv-\Gamma_\text{in}\left(\mathcal{G}\right)\right)$ and $C$ is the local update of $\Gamma_\text{in}\left(\mathcal{G}\right)$. 
However, in order to calculate this determinant, we need to calculate $M$'s inverse. 
For that, we use the Woodbury formula which computes $M$'s determinant after the update in terms of its determinant before the update,
\begin{equation}
  (M+UCV)\inv=M\inv-M\inv U(C\inv+VM\inv U)\inv V M\inv.
  \label{eq:woodbury}
\end{equation}
Thus, at each step, we can update the determinant and the inverse according to the previous inverse, by computing the inverse of the update $C$. 
This allows computation of the weights when computing the inverses and determinants from scratch only once --- at the first MC step. 
For local updates, where $C$ has fixed dimensions, the computational complexity is significantly reduced to $\mathcal{O}(N^2)$ instead of $\mathcal{O}(N^3)$ (for a matrix of size $N\times N$), as it only requires computing the inverse and determinant of a matrix much smaller than $M$. 

\begin{figure*}
\includegraphics{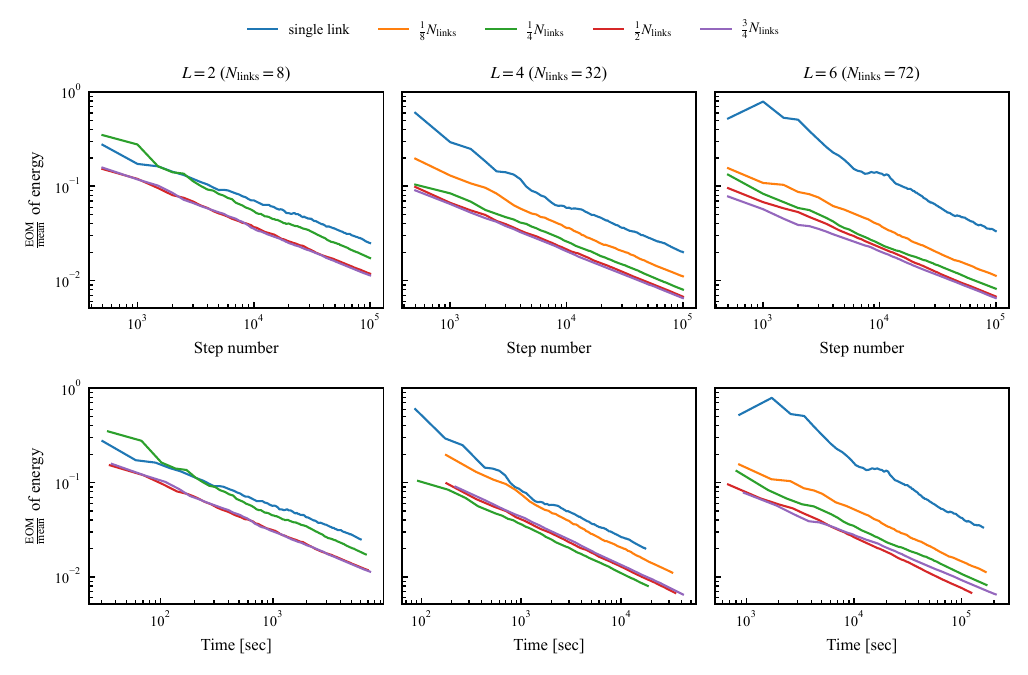}
    \caption{Relative error on the mean of the energy as a function of step number and time for different number of updated links per step and various lattice size $L \times L$. 
    All evaluations were performed for Hamiltonian couplings $g_{\text{int}} = 1$, $g_{\text{mass}} = 1$, and $g_\text{E} = \frac{1}{4g_\text{B}} = 1.25$.
    In terms of step number, the EOM typically converges faster when more links are updated. 
    However, since larger updates are more computationally expensive, there is a trade-off. 
    The fastest convergence in terms of computational time is usually achieved when fixing about $\frac{1}{4}$ to $\frac{1}{2}$ of the links. 
    For clarity, results with more than $\frac{3}{4}N_\text{links}$ updated links are omitted, as the corresponding curves were nearly indistinguishable.}
    \label{fig:eom_update_size}
\end{figure*}

We find that, when performance is evaluated in terms of computational time rather than step number, the most efficient choice is typically to update about $\frac{1}{4}$ to $\frac{1}{2}$ of the links per step (see, e.g., Fig.~\ref{fig:eom_update_size}). This reflects a balance between two opposing effects: updating more links per step reduces autocorrelation and accelerates the decay of the EOM, but doing so also increases the computational cost due to the increased number of updates done per measurement. 
This increase in runtime is further amplified by the probability of a non-trivial change in the gauge configuration between measurements: in cases where the Monte Carlo proposal is rejected, we can reuse previously calculations, rather than computing values from scratch. However, a larger update size increases the probability that at least one link will flip, leading to a higher frequency of computationally expensive, non-cached evaluations. This effect is particularly significant for small discrete gauge groups. For instance, in the $\mathbb{Z}_2$ theory simulated here, a single random proposal for the update has a $50\%$ chance of resulting in the same configuration, which allows the simulation to utilize cached values for efficient measurement, but this drops to only $0.5^8$ for an update size of 8. This huge disparity is slightly reduced by accounting for proposal acceptance, but remains significant.

\begin{figure}
\includegraphics{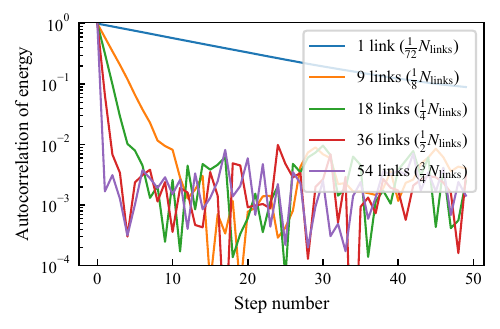} 
    \caption{ Autocorrelation of the energy as a function of step number for different number of updated links per step. 
    All evaluations were performed for a $6\times 6$ lattice with Hamiltonian couplings $g_{\text{int}} = 1$, $g_{\text{mass}} = 1$, and $g_\text{E} = \frac{1}{4g_\text{B}} = 1.25$. 
    Typically, the more links modified per step number, the faster the autocorrelation decays. 
    }
    \label{fig:binning_update_size}
\end{figure}

Our conclusion follows from studying how different choices for the number of updated links affect the convergence behavior of the MC evaluations. When more than one link is updated in a single step, the local update scheme is applied sequentially. As expected, increasing the number of updated links generally reduces autocorrelation and causes the EOM to decrease more rapidly as a function of step number, as seen in typical MC runs shown in Figs.~\ref{fig:binning_update_size} and~\ref{fig:eom_update_size}. These observations illustrate the trade-off underlying the optimal choice of update size.

\section{Gauge Fixing} \label{sec:gauge_fixing}
When computing expectation values of physical observables, we integrate over all gauge field configurations, as depicted in Eq.~\eqref{eq:general_exp_val}. 
However, since both the ansatz and the observables of interest are gauge invariant, we can fix the gauge over trees on the lattice and integrate only over the remaining links to reduce the number of configurations involved in the integration. 
For exact contraction (EC) i.e., exact evaluation of the sum in Eq.~\eqref{eq:general_exp_val} for discrete groups, this reduction directly improves computational efficiency by reducing the number of terms in the sum. 
In contrast, for MC evaluations, the impact of gauge fixing is less straightforward. 
Previous studies, including conventional imaginary time  action-based LGT Monte Carlo simulations~\cite{VLADIKAS198693,LAUTRUP1987188}, have observed that gauge fixing can lead to slower thermalization and error convergence. 
Here, we investigate how fixing the gauge for various cases affects error convergence in MC computations of GGFPEPS.

We start by discussing why fixing the gauge and sampling only non-fixed links is possible --- as we shall see now, the integrand in Eq.~\eqref{eq:general_exp_val}, $F_\mathcal{O}(\mathcal{G})p(\mathcal{G})$, is pure gauge invariant, i.e., invariant under transformations including only the gauge field degrees of freedom.
A pure gauge transformation for a gauge field $g(\site{x},i)$ on a link $(\site{x},i)$ is of the form, 
\begin{equation}
    g(\site{x},k)\to \Tilde{g}(\site{x},i)=h\inv(\site{x})g(\site{x},i)h(\site{x}+\hat{\mathbf{e}}_i),
\label{eq:gauge_transformation}
\end{equation} 
where $h(\site{x})$ are group elements associated with the sites. 
We would like to use our ansatz gauge invariance to fix links over trees on the lattice to a fixed value, e.g., the identity group element $g=e$. 
Gauge fixing is always possible on trees due to their loop-free structure as shown in appendix \ref{sec:gf_trees}.   

Under a pure gauge transformation $\mathcal{G}\to \Tilde{\mathcal{G}}$, the fermionic part of the ansatz transforms unitarily taking the form (see appendix \ref{sec:gauge_inv_p}), 
\begin{equation}
\ket{\psi(\Tilde{\mathcal{G}})}=\prod_{\boldsymbol{y}\in\text{odd}} \Tilde{\theta}_{h\left(\boldsymbol{y}\right)}\left(\boldsymbol{y}\right)\prod_{\boldsymbol{x}\in\text{even}} \theta^{\dagger}_{h\left(\boldsymbol{x}\right)}\left(\boldsymbol{x}\right)\ket{\psi(\mathcal{G})}\equiv \hat{\mathcal{U}}\ket{\psi(\mathcal{G})}.
\end{equation}
This implies that the probability distribution $p_0\left(\Tilde{\mathcal{G}}\right)$, remains invariant
\begin{equation}
p_0\left(\Tilde{\mathcal{G}}\right)=\braket{\psi\left(\Tilde{\mathcal{G}}\right)}  
=\bra{\psi\left(\mathcal{G}\right)}\hat{\mathcal{U}}^{\dagger}\hat{\mathcal{U}}\ket{\psi\left(\mathcal{G}\right)}=p_0\left(\mathcal{G}\right).
\end{equation}

All that remains is to show that the function $F_\mathcal{O}(\mathcal{G})$ is pure gauge invariant. 
For observables $O$ that depend solely on the gauge configuration, such as the electric energy or Wilson loops, this is straightforward, as they are functions only of the gauge field.

Also the function $F_\mathcal{O}(\mathcal{G})$ of expressions involving fermionic operators can be pure gauge invariant.
Operators of interest are mesonic operators $\mathcal{M}_f\left(\site{x},\mathcal{C},\site{y}\right)$, which are defined along a path $\mathcal{C}$ connecting the site $\site{x}$ to site $\site{y}$,
\begin{equation}
\mathcal{M}_f\left(\site{x},\mathcal{C},\site{y}\right) = 
\psi^{\dagger}_{m}\left(\site{x}\right)
    \left[\underset{\ell \in \mathcal{C}}{\prod}U\left(\ell\right)\right]_{mn}
    \psi _{n}\left(\site{y}\right).
    \label{mesdef}
\end{equation}
These operators are more complex because they depend on both the matter and the gauge field configuration, rather than solely on the gauge field. 
Their $F_{\mathcal{M}_f}(\mathcal{G})$ function is also composed of matter and gauge fields terms,
\begin{equation}
\begin{aligned}
F_{\mathcal{M}_f}\left(\mathcal{G}\right) ={}& 
\left[\underset{(\ell)\in\mathcal{C}}{\prod}
    D\!\left(g(\ell)\right)\right]_{mn} \\
&\times
\frac{\bra{\psi(\mathcal{G})}
      \psi_{m}^{\dagger}(\site{x})\psi_{n}(\site{y})
      \ket{\psi(\mathcal{G})}}
     {\braket{\psi(\mathcal{G})|\psi(\mathcal{G})}}
\end{aligned}
\end{equation}
As a result, it is not trivially pure gauge invariant. 
Nevertheless, the transformation of the gauge field part of $\left[{\prod}D\left(g\left(\ell\right)\right)\right]_{mn}$ is exactly compensated with the matter state $\ket{\psi(\mathcal{G})}$ transformation, $\hat{\mathcal{U}}$, acting on the fermionic operators $\psi^\dagger$. 

Therefore, we conclude that, as expected, for all gauge invariant operators,
$F_\mathcal{O}(\mathcal{G})=F_\mathcal{O}(\mathcal{\Tilde{G}})$. 
Thus the integrand in Eq.~\eqref{eq:general_exp_val} could be replaced, without changing the integration variable,
\begin{equation}
    \left\langle \mathcal{O} \right\rangle = \int \mathcal{DG} \ F_{\mathcal{O}}\left(\mathcal{G}\right)p\left(\mathcal{G}\right) = \int \mathcal{DG} \ F_{\mathcal{O}}\left(\Tilde{\mathcal{G}}\right)p\left(\Tilde{\mathcal{G}}\right).
\end{equation}
For each gauge configuration $\mathcal{G}$, we can choose a $\Tilde{\mathcal{G}}$ in which the links over some tree are all fixed (e.g., to the identity group element $e$). 
As a result, the integration over these links becomes trivial, factorizing out a constant that is insignificant due to normalization. 
Therefore, we conclude that the gauge can be fixed over trees, and the integration can be performed over the non-fixed links only. 
In the following, we study the effect of this transformation on the convergence properties of the MC simulation.

Gauge fixing can be implemented both in MC and exact contraction (EC) schemes.
In EC it reduces the complexity of evaluating physical observables significantly. 
In fact, for a finite gauge group $G$ of order $\left|G\right|$, we achieve roughly a square root improvement in the computational scaling of the EC runs -- from $O(\left|G\right|^{N_{\text{links}}})$ to $O(\left|G\right|^{\frac{N_{\text{links}}}{2}+1})$ for a $2+1$ dimensional maximal tree. 
This improvement makes it possible to compute the ground state using EC for $4\times 4$ $\mathbb{Z}_2$ LGTs, which was previously infeasible.
Furthermore, it can be particularly useful for future developments of the code, where we aim to extend it to larger groups or higher dimensions, for which EC would otherwise be infeasible without gauge fixing. 
Moreover, the EC speedup can be valuable for comparing with MC results as well as for debugging and implementation purposes.

\begin{figure}
\includegraphics[width=0.95\linewidth]{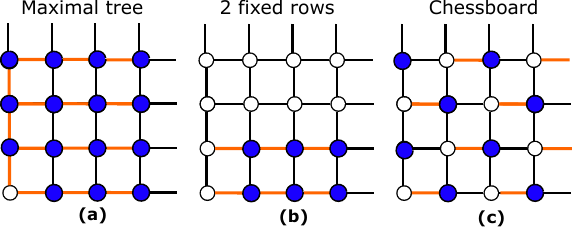}
    \caption{The types of gauge fixing trees we examined. Blue sites represent sites where the $h$ values are fixed according to Eq.~\eqref{eq:gauge_transformation}. 
    Orange links indicate links fixed to the identity value these links are included in the tree. 
    Panel (a) describes a maximal tree where all horizontal links in each row except the last one, and all vertical links in a single column except the last one are fixed. 
    Panel (b) describes a tree where two rows are fixed. Panel (c) describes \enquote{chessboard} tree, that includes all horizontal links $\ell=(\site{x},1)$ coming out of even sites $\site{x}$.}
    \label{fig:tested_trees}
\end{figure}

\begin{figure*}
    \centering
    \includegraphics[width=0.95\textwidth]{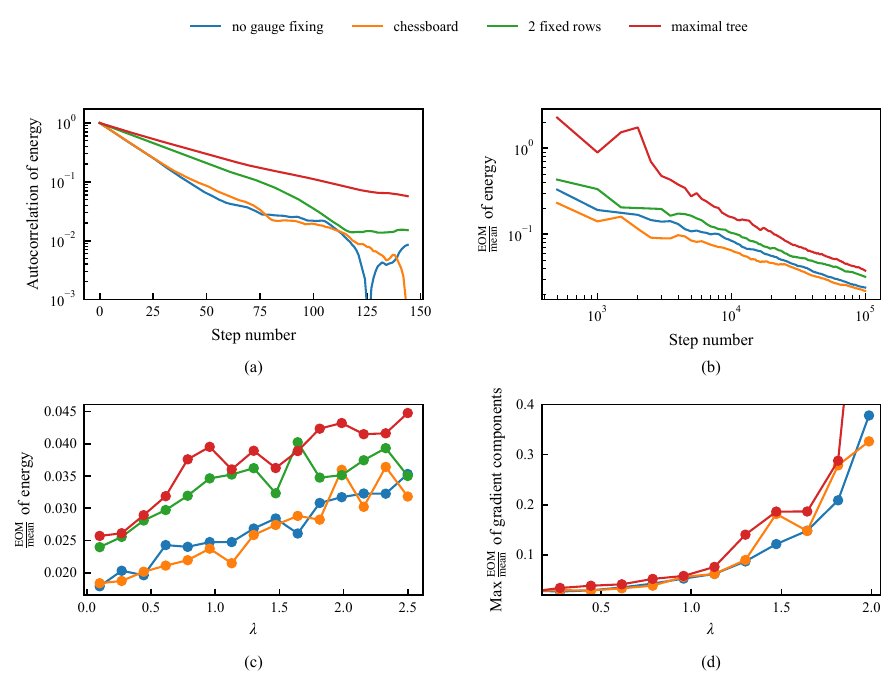}
    \caption{Effect of different gauge fixing trees on MC evaluation metrics for a $6 \times 6$ lattice with Hamiltonian couplings $g_{\text{int}} = 1$ and $g_{\text{mass}} = 1$.
    (a) Autocorrelation and (b) relative error on the mean (EOM) of the energy as a function of step number for $g_E = \frac{1}{4g_B} = 0.39$. \enquote{No gauge fixing} and \enquote{chessboard} trees exhibit the fastest autocorrelation decay and EOM convergence, with \enquote{chessboard} being slightly better. Fixing two rows is noticeably worse, and the maximal tree performs worst of all. For step numbers larger than the typical decay time $\tau$, the autocorrelation begins fluctuating between $10^{-2}$ to $10^{-3}$.
    (c) Relative EOM of the energy and (d) maximal relative EOM among energy gradient components after $10^5$ MC steps as a function of Hamiltonian's coupling $\lambda=2g_E=\frac{1}{2g_B}$. For nearly all $\lambda$ values, \enquote{no gauge fixing} and \enquote{chessboard} exhibit lower EOM than \enquote{maximal}, with the trend becoming more distinct for $\lambda > 1$.
    \textit{General remarks:} Trees with other numbers of fixed rows are omitted for clarity; their values consistently lie between those of the chessboard and maximal trees. In panel (d), the regime $\lambda > 2$ and the last data point of \enquote{maximal tree} for $\lambda=2$ were omitted for clarity due to a much higher EOM to mean ratio, though the overall trend persists in this regime with larger deviations.}
    \label{fig:gf_combined}
\end{figure*}

We would like to see whether this improvement carries over to MC evaluations. 
To study the effect of gauge fixing on the convergence rate of MC evaluations, we analyzed the autocorrelation and error on the mean (see Fig.~\ref{fig:gf_combined}) of various observables as a function of step number, considering different Hamiltonian couplings, lattice sizes, ansatz parameters and gauge-fixing trees. 
Specifically, we examine three types of trees (and disconnected trees) --- a maximal tree that includes all horizontal links in each row except the last one, and all vertical links in a single column except the last one, trees with a fixed number of rows and a \enquote{chessboard} tree, that includes all horizontal links $\ell=(\site{x},1)$ coming out of even sites $\site{x}$ (see Fig.~\ref{fig:tested_trees}).

In about $64\%$ of the tested cases, the mean values of the total energy obtained using different gauge fixing trees (including the unfixed case), agree within a single EOM margin, based on simulations with $10^5$ warmup steps and $10^5$ measurement steps. 
In about $95\%$ of the cases, the deviation remains within two EOM margins. 
In addition, the acceptance ratio, which is the mean of accepting an update, varies within $2\%$ for different trees. 
Both of these measurements agree with what we expect from probability theory and demonstrate good error calibration of our algorithm.

As seen in typical MC evaluations in Fig.~\ref{fig:gf_combined}, for almost all cases, no gauge fixing and \enquote{chessboard}
performed best --- lowest autocorrelation decay time and fastest EOM convergence --- while the maximal tree performed worst. 
Different number of fixed rows performed worse than no gauge fixing and chessboard, but significantly better than maximal tree. 
In most cases the \enquote{chessboard} tree exhibited an EOM and autocorrelation decay rate comparable to the unfixed case (typically a slightly faster EOM and a slightly slower autocorrelation), while remaining significantly faster than configurations with fixed rows and maximal tree. 
A similar trend is observed for the energy gradient as well, as shown in Fig.~\ref{fig:gf_combined}(d). 
This is a key observable, as it controls the variational minimization of the ansatz.

\begin{figure}[htb]
\includegraphics[width=\linewidth]{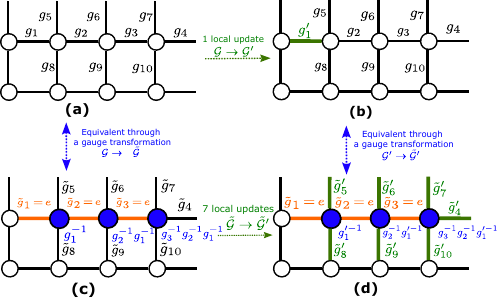}
    \caption{Demonstration of the impact of gauge fixing on the number of local updates required in a $4$ columns lattice. 
    Panels (a) and (b) depict a single local update in the non-gauge-fixed setting. 
    Panels (c) and (d) show the corresponding configurations after gauge fixing, where a single row has been fixed. 
    Blue sites represent sites where the $h$ values are fixed according to Eq.~\eqref{eq:gauge_transformation}. 
    Orange links indicate links fixed to the identity value $e$ due to gauge fixing, while green links mark links that are modified during the update. 
    As seen in the gauge fixed case ((c) and (d)), performing what is locally a single update in the unfixed setting now requires a coordinated, global update involving 7 links.} 
    \label{fig:updates_when_gauge_fixing}
\end{figure}

A possible explanation is that, for connected gauge fixing trees (such as maximal trees or entire rows), a single local update in the non-fixed configuration can correspond to a complex, non-local change in the gauge-fixed configuration, requiring many update steps to achieve the same effect. 
For example, consider Fig.~\ref{fig:updates_when_gauge_fixing}. 
Fixing the gauge fields along an entire row in a lattice with $4$ columns transforms a local MC update (easily performed in the non-gauge-fixed setting) into a global update in the gauge-fixed settings, requiring modifications to $7$ links. 
In general, when performing an update on a link (in the non-gauge-fixed setting) when that link is to be fixed, the number of links required for an update (in the gauge fixed setting) scales linearly with the row length. Thus, some nearby configurations in the non-gauge fixed case are transformed into distant configurations, affecting the distribution of sampled configurations.
In contrast, gauge fixing of disconnected links, such as in a  \enquote{chessboard} pattern, where fixed links are spatially separated and not directly connected, reduces this issue. 
In such a setup, modifying one of the corresponding links (which is fixed in the gauge-fixed setting) can still be achieved through a small number of local updates --- a maximum of just $3$. 

\section{Translation invariance} \label{sec:TI}
The ansatz is constructed to be translationally invariant with respect to one site shifts. 
Furthermore, the pure gauge part of the Hamiltonian, which contains the electric and magnetic energies, is translationally invariant. 
This provides us with two different approaches for computing these expectation values of these term --- we can either use that the observables are translation invariant and compute them for one link or plaquette, multiplying by the number of links or plaquettes, or compute them explicitly for all the links. 
The latter has the advantage that each step already incorporates an average over several links across the lattice. 
However, it increases the computation time for an MC measurement step. 
In this section, we study this trade-off for both the magnetic and electric energy terms.

\begin{figure}[htb]
\includegraphics{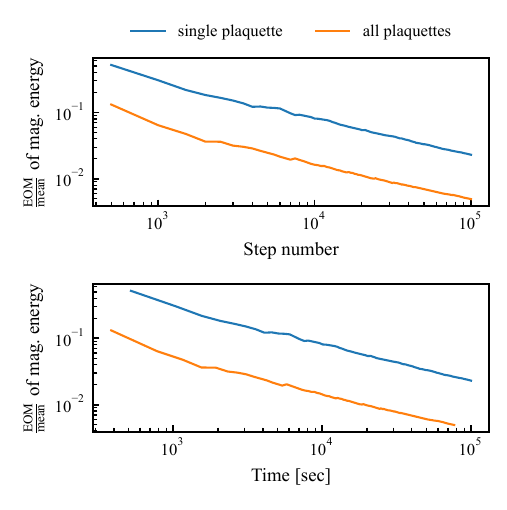}
    \caption{Error on the mean over mean of magnetic energy as a function of step number and computation time when sampling a single plaquette vs. when sampling all plaquttes. 
    This evaluation was performed for a $6$ by $6$ lattice with Hamiltonian couplinds $g_\text{int}=1$, $g_\text{mass}=1$ and $g_\text{E}=\frac{1}{4g_\text{B}}=0.39$. 
    It represents a typical MC evaluation.
    The magnetic energy error converges much faster when averaging over all plaquettes in terms of both time and step number.}
    \label{fig:trans_inv_mag_eom}
\end{figure}

For the magnetic energy, extending the computation to all plaquettes requires only a straightforward generalization --- iterating over all plaquettes and computing the corresponding Wilson loops. 
Since the computation of the MC observable function $F_\mathcal{O}(\mathcal{G})$ is performed in the magnetic basis, it could be implemented quite easily, and does not extend the computational time significantly. 
Therefore, as can be seen in Fig.~\ref{fig:trans_inv_mag_eom}, averaging over all plaquettes yields much better magnetic energy EOM convergence than the single plaquette, with respect to both step number and wall-clock time. 

\begin{figure}[htb]
\includegraphics{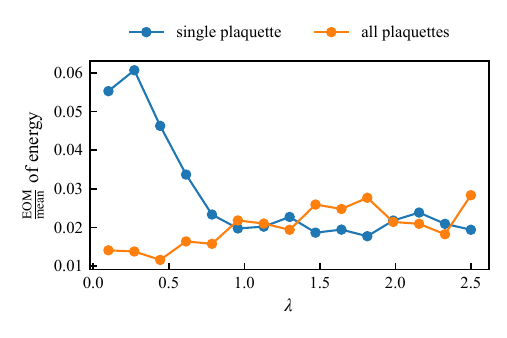}
    \caption{Error on the mean over mean of the energy after $10^5$ MC steps as a function of the Hamiltonian coupling $\lambda=2g_E=\frac{1}{2g_B}$, comparing cases where the magnetic energy is averaged over all plaquettes versus a single plaquette. 
    All evaluations were performed on a $6\times6$ lattice with $g_\text{int}=1$ and $g_\text{mass}=1$. 
    For small $\lambda$ (corresponding to large $g_B$), the EOM is significantly higher when averaging over all plaquettes, whereas for larger $\lambda$, where the magnetic energy is less dominant, the EOM values of the two cases become comparable.} 
    \label{fig:trans_inv_mag_eom_to_g}
\end{figure}

The situation is more complex for the total energy, as seen in Fig.~\ref{fig:trans_inv_mag_eom_to_g}. 
Including additional energy terms, such as the electric energy, introduces additional \enquote{noise}.
Depending on the coupling strength, it can be significant enough to offset the EOM reduction gained by spatially averaging the magnetic energy.
Consequently, for the total energy, the EOM convergence rate depends on the value of $\lambda = 2g_\text{E}=\frac{1}{2g_\text{B}}$. 
For small $\lambda$ (large $g_\text{B}$) values, the EOM of the energy typically converges faster when averaging all plaquettes. 
However, for large $\lambda$ (small $g_\text{B}$) values, the EOM convergence rate was comparable between the two methods with the single plaquette sampling showing a slightly faster rate. 
This trend, however, was not entirely consistent and showed some variation depending on the specific ansatz parameters.

To extend the improvement to larger coupling parameters, it is desirable to also average over the electric energy in a similar fashion.
However, the computation of the electric energy $F_\mathcal{O}(\mathcal{G})$ is not as straightforward since it is also performed in the magnetic basis. 
Unlike the magnetic energy, the electric energy is not diagonal in this basis.
To calculate the electric energy on a single link, we compute a modified covariance matrix that treats the virtual modes on that link as physical. 
Then, in order to get the electric energy on that link, we perform operations on that covariance matrix, such as computing Pfaffians. 
Performing this for several links introduces an overhead linear in the number of links on which we calculate the electric energy. 
Unlike the magnetic energy, where we simply access memory to retrieve the gauge group elements on each plaquette and multiply them over all plaquettes, the operations required for each link here are non-negligible in terms of computational cost. 
Therefore, computing the electric energy for several links will extend computation times.

\begin{figure}[htb]
\includegraphics{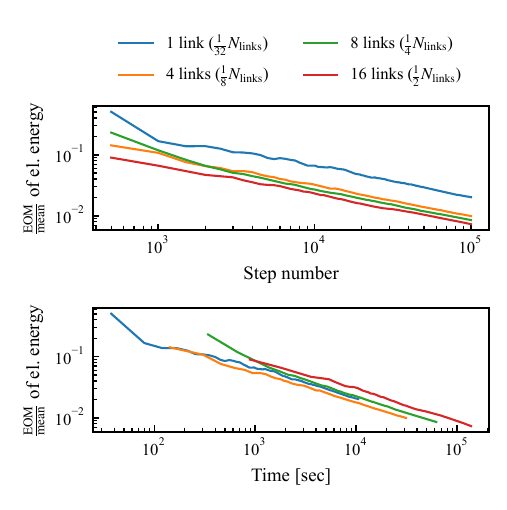}
    \caption{Error on the mean over mean for the electric energy as a function of step number and time for various numbers of links over which the electric energy is averaged.This evaluation was performed for a $4\times4$ lattice with Hamiltonian's couplings $g_\text{int}=1$, $g_\text{mass}=1$ and $g_\text{E}=\frac{1}{4g_\text{B}}=1.25$.
    With respect to step number, fastest convergence is achieved for $16$ links -- the more links averaged over, the faster the convergence. 
    However, with respect to time, averaging over more links typically slows down the convergence and fastest is observed for $1$ or $8$ links.}
    \label{fig:trans_inv_el_time}
\end{figure}

\begin{figure}
\includegraphics{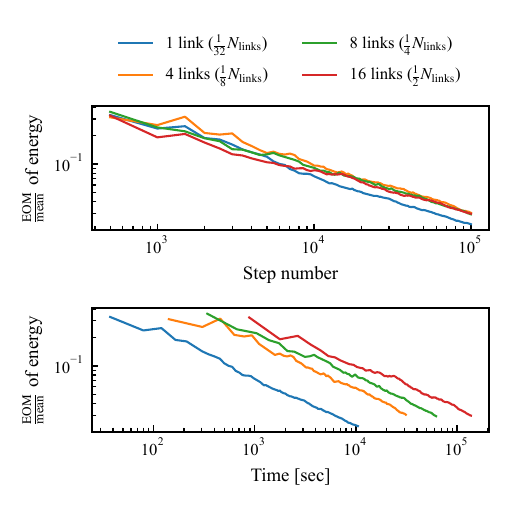}
    \caption{
    Error on the mean over mean for the total energy as a function of step number and time for various numbers of links over which the electric energy is averaged. 
    This evaluation was performed for a $4\times4$ lattice with Hamiltonian's couplings $g_\text{int}=1$, $g_\text{mass}=1$ and $g_\text{E}=\frac{1}{4g_\text{B}}=1.25$.
    With respect to both step number and time, fastest convergence is achieved for a single link. 
    While the convergence with respect to the step number was not entirely consistent across different settings, the convergence with respect to time consistently showed the fastest decay for the single-link case. 
}
    \label{fig:trans_inv_el_time_total_energy}
\end{figure}

As shown in Fig.~\ref{fig:trans_inv_el_time}, in terms of step number, increasing the number of links averaged over improves the convergence rate of the electric energy error. 
However, when considering computation time, for a $4\times4$ lattice, the fastest convergence is achieved for either a single link or $\frac{1}{8}N_\text{links}$. 
For the total energy, the fastest convergence with respect to time is obtained for a single link (see Fig.~\ref{fig:trans_inv_el_time_total_energy}). 
This implies that in the current implementation, although averaging over all links in the lattice enhances the EOM convergence per step, the additional computational cost per step overcomes the benefit, making it preferable to compute the energy over only a single link. 

\section{Conclusions and Outlook} \label{sec:conclusions}
The usefulness of numerical methods crucially depends on their efficiency. In this work, we present numerical improvement to the GGFPEPS method, a variational Monte Calro optimization of a tensor network construction for lattice gauge theories.
In our analysis, we focus on two features: gauge fixing to reduce the number of independent gauge field variables and the translational invariance of the ansatz.

The gauge fixing procedure developed in this work yields prescription for fixing the gauge fields over trees in GGFPEPS, which reduces the dimensionality of the integration. 
For discrete groups in $2+1$ dimensions, this procedure decreases the number of gauge configurations by up to nearly a square-root factor. 
When the ansatz is contracted exactly, i.e., when Eq.~\eqref{eq:general_exp_val} is evaluated explicitly, the computational efficiency improves significantly, since we avoid a sizable part of the exponentially scaling number of configuration. 
This improvement makes it feasible to run $4\times4$ $\mathbb{Z}_2$ LGT variational ground state search using GGFPEPS without resorting to Monte Carlo.
This development also enables exact contractions for more complex groups. An essential feature to benchmark the ansatz independent of the Monte Carlo convergence. 
While highly beneficial for exact contractions, we found that fixing the gauge reduced MC convergence rate in most cases. 
We attribute this to the fact that a local link update in the unfixed settings corresponds to a highly non-local update after gauge fixing, involving a number of links proportional to the length of the fixed string. 
This interpretation is supported by the observation that chessboard tree, -- in which the gauge is fixed over unconnected trees each containing only a single link -- performed comparably to the unfixed case.

We examined the difference between space- and time-averaging in GGPEPS simulations. By exploiting the translational invariance, we can significantly speed up each measurement step. This strategy, however, comes at the expense of not using the different links in the lattice as an additional averaging opportunity. 
When averaging explicitly over all plaquettes of the system, the convergence rate of the magnetic energy improves.
This improvement carries over to the total energy for an extensive coupling range at high coupling.
At low coupling, however, the electric energy dominates and averaging the electric energy is computationally more expensive since sampling is performed in the magnetic basis.
This suggests that in future simulations, the magnetic energy should be computed by averaging over all plaquettes, while the electric energy will be evaluated on a single link. 
Additionally, we found that increasing the number of varied links per step improves convergence with respect to step number. The trade-off between error reduction and computational cost favors updates of a quarter to half of the lattice links.

The research presented in this paper can be naturally extended in several directions.
Since the gauge fixing procedure extends to non-Abelian group, GGPEPS simulations for these groups without Monte Carlo sampling become feasible. This allows us to examine the quality of the ansatz independent of sampling errors.
By actively exploiting space averaging, the number of samples can be reduced to enable the simulation of larger lattices, possibly even transition to simulations in three spatial dimensions.

\section*{Data Availability}
All data presented above is available in a Zenodo repository at~\cite{gomelski_2026_19351766}. The repository also
includes a script that generates all the plots.

\acknowledgements
We would like to thank I. Cirac, G. Roose, E. Itou,A. Yosprakob, and T. Yoshida for fruitful discussions. 

This research is funded  by the European Union
(ERC, OverSign, 101122583). 
P.E. acknowledges the support received through the NWO-Quantum Technology programme (Grant No.~NGF.1623.23.006)
and funding by the Carl-Zeiss-Stiftung (CZS Center QPhoton).
Views and opinions expressed are those of the authors only and do not necessarily reflect those of the European Union or the European Research Council.

The results were computed using computing resources at the Fritz Haber Center for Molecular Dynamics at The Hebrew University of Jerusalem.
\appendix
\section{Gauge fixing over trees} \label{sec:gf_trees}
To demonstrate how we can fix the gauge over trees, let us consider a square lattice of size $L\times L$ where each link $\left(\boldsymbol{x},k\right)$ contains some group element $g\left(\boldsymbol{x},k\right)$.
Here, we will focus on pure gauge transformations, which act solely on the gauge field and do not affect the matter. These transformation are of the form of Eq.~\eqref{eq:gauge_transformation},
\begin{equation} \label{eq:gauge_transformation_2}
g\left(\site{x},i\right)\rightarrow \Tilde{g}\left(\site{x},i\right)=h^{-1}\left(\site{x}\right)g\left(\site{x},i\right)h\left(\site{x}+\boldsymbol{\hat{e}}_{i}\right)
\end{equation}
where $h(\site{x})$ are group elements associated to the sites. 

Before dealing with a general tree, we shall first demonstrate how to fix $\Tilde{g}\left(\site{x}=(x_1,x_2),i\right)$ on a given row $x_2=const$, i.e. a path $\left\{g\left((x_1,x_2),1\right)\right\}_{x_1=0}^{L-1}$. This would allow us to develop intuition and subsequently extend this path to a general tree structure. Imposing $\Tilde{g}=e$ along this path yields a recursion relation for $h$
\begin{equation} 
h^{-1}\left(x_1,x_2\right)g\left((x_1,x_2),1\right)h\left(x_1+1,x_2\right)=e.
\end{equation}
For $x_1>0$ the solution is of the form
\begin{align}
h(x_{1},x_{2})
&= g^{-1}\!\big((x_{1}-1,x_{2}),1\big)\cdots g^{-1}\!\big((1,x_{2}),1\big) \nonumber\\
&\qquad\times g^{-1}\!\big((0,x_{2}),1\big)\, h(0,x_{2}). \label{eq:sol_rec}
\end{align}
and due to the open boundary condition, $h\left(0,x_2\right)$ can be arbitrarily set. Note that fixing $h$ in this manner also impacts adjacent links along the path. 

This procedure can be repeated for any $x_2$ and thus fixing half of the links to the identity. In addition due to the freedom to choose $h\left(0,x_2\right)$, we can also set the first column to the identity element, i.e. $g\left(\left(0,x_2\right),2\right)=e$ in a similar manner to (\ref{eq:sol_rec}). For periodic boundary conditions, we cannot fix the last links in each row due to the extra constraint  $h\left(0,x_2\right)=h\left(L-1,x_2\right)$. Furthermore, we cannot fix the last  link in the first column. Nevertheless, even in this case it is possible to fix roughly half of the links in the lattice.

This provides a recipe to fix the gauge over a specific maximal tree, which is sufficient for our purposes. However, we would like to demonstrate that it can be generalized to any chosen tree. Furthermore, we will show that in general the gauge field cannot be fixed over closed loops, which implies that a maximal tree contains the maximal amount of links that can be fixed. The solution for $h(\site{x})$ in (\ref{eq:sol_rec}) can be conceptualized as a telescoping product, originating from $h\left(0,x_2\right)$ and successively multiplied with $g^{-1}$ from the left along a path towards the site $\boldsymbol{x}=\left(L-1,x_2\right)$. This intuition lays the groundwork for extending this method to a general tree. 

Let us consider an arbitrary tree. We start by selecting a specific site as the origin, denoted as $O$ and set $h(O)=e$. Next, we define paths along the tree, starting from the origin and extending to the edges of the tree. When choosing these paths, we only need to ensure that each link has at most one path leading to it from the origin, which is possible due to the tree structure.

\begin{figure}
    \centering
    \includegraphics[width=0.75\linewidth]{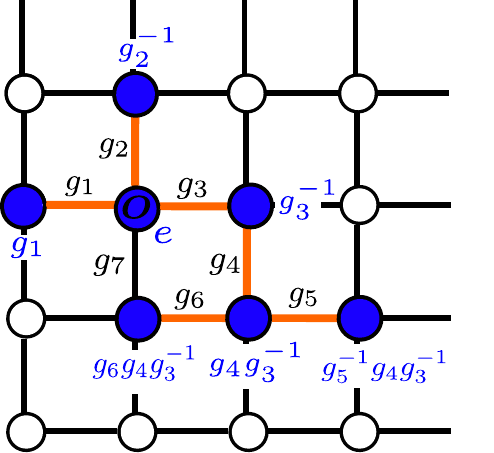}
    \caption{Demonstration of the gauge fixing procedure results. Sites where the function $h$ is set appear in blue, where $h$'s value is shown around them in blue. The origin is denoted as in the main text by $O$. It can be verified that by a gauge transformation of the form of (\ref{eq:gauge_transformation}) all orange links are gauge fixed to the identity group element. Note that the link $g_7$ cannot be fixed because it closes a loop meaning there is no additional $h$ on its edges that can be set freely.}
    \label{fig:gauge_fixing_demo}
\end{figure}

Next, we go along these paths and set $h$ in the following manner:
\begin{enumerate}
\item When moving along the link $g\left(\site{x},i\right)$ in the positive direction, $+\boldsymbol{\hat{e}}_{i}$, we set
\begin{equation}
h\left(\site{x}+\boldsymbol{\hat{e}}_{i}\right)=g^{-1}\left(\site{x},i\right)h\left(\site{x}\right).
\end{equation}

\item When moving along the same link in the negative direction, $-\boldsymbol{\hat{e}}_{i}$, we set
\begin{equation}
h\left(\site{x}-\boldsymbol{\hat{e}}_{i}\right)=g\left(\site{x},i\right)h\left(\site{x}\right).
\end{equation}
\end{enumerate}
This process is demonstrated in Fig.~\ref{fig:gauge_fixing_demo}. 

This recursive process ensures that all links along the tree are eventually fixed to the identity. The specific structure of the tree ensures that we can choose paths in such a way that no two different paths arrive at the same link, and each path contains each link at most once. In particular, it is impossible to fix closed loops because doing so would require paths to affect the same link more than once. Equivalently, there would exist some site where $h$ is set more than once. For instance consider the link $g_7$ in Fig.~\ref{fig:gauge_fixing_demo} which cannot be fixed because its two neighboring sites have already been set.
This implies that a maximal tree contains the maximal amount of links that can be fixed.

\section{Gauge invariance of $p(\mathcal{G})$} \label{sec:gauge_inv_p}

Let us show that $p(\mathcal{G})$ is pure gauge invariant. For that, it is enough proving that $p_0(\mathcal{G})\equiv\braket{\psi\left(\mathcal{G}\right)}$ is pure gauge invariant. Let us consider some gauge configuration $\Tilde{\mathcal{G}}$ which is equivalent to $\mathcal{G}$ through some gauge transformation such as \eqref{eq:gauge_transformation}. Under gauge transformations, the state  $\ket{\psi\left(\mathcal{G}\right)}$ changes only by the action of the operator $\mathcal{U}_g\left(\site{x},i\right)$, which transforms in the following way:
\[
\ensuremath{\mathcal{U}_{g}\left(\site{x},i\right)=\begin{cases}
\Tilde{\theta}_{g} & \site{x}\ \text{even}\\
\theta_{g}^{\dagger} & \site{x}\ \text{odd}
\end{cases}\rightarrow\begin{cases}
\Tilde{\theta}_{h^{-1}\left(\site{x}\right)\,g\,h\left(\site{x}+\boldsymbol{\hat{e}}_{i}\right)} & \site{x}\ \text{even}\\
\theta_{h^{-1}\left(\site{x}\right)\,g\,h\left(\site{x}+\boldsymbol{\hat{e}}_{i}\right)}^{\dagger} & \site{x}\ \text{odd}
\end{cases}}
\]
\begin{equation}
    =\begin{cases}
\Tilde{\theta}_{h\left(\site{x}+\boldsymbol{\hat{e}}_{i}\right)}\Tilde{\theta}_{g}\Tilde{\theta}_{h\left(\site{x}\right)}^{\dagger} & \site{x}\ \text{even}\\
\theta_{h\left(\site{x}+\boldsymbol{\hat{e}}_{i}\right)}^\dagger\theta_{g}^{\dagger}\theta_{h\left(\site{x}\right)} & \site{x}\ \text{odd}
\end{cases},
\end{equation}
where all operators act on the $(\boldsymbol{x},k)$ link. 

\begin{figure*}
\includegraphics[width=0.9\linewidth]{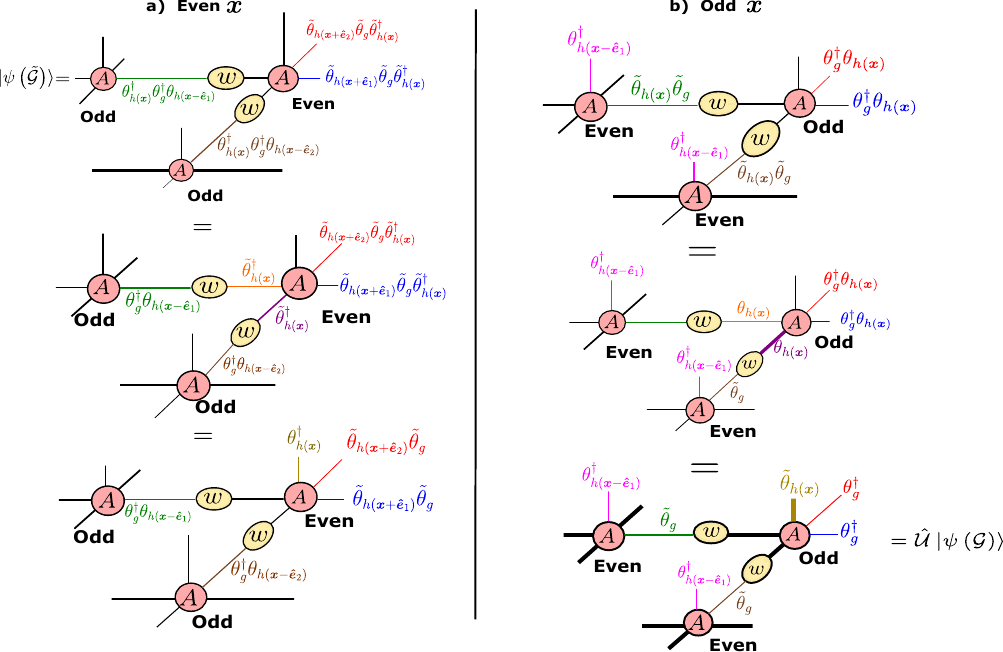}
\captionsetup[subfigure]{labelformat=empty}
    \caption{Illustration of the transformation of $\ket{\psi({\mathcal{G}})}$ under a gauge transformation as defined by Eq.~\eqref{eq:gauge_transformation}. The colors
    indicate which link each operator acts on. The transformation for the odd sites is depicted after the transformation for the
    even sites has been applied. All steps are based on the rules provided in~\cite{kelman2024gauged}. This gauge transformation results in a unitary transformation $\hat{\mathcal{U}}$ avting on the matter, as shown in Eq.~\eqref{eq:matter_transformation_wave_func}.}
   \label{fig:p_G_proof}
\end{figure*}

Building on this concept, Fig.~\ref{fig:p_G_proof} demonstrates that \(\ket{\psi(\mathcal{G})}\) transforms to \(\ket{\psi(\mathcal{\Tilde{G}})}\)  under a unitary operation $\hat{\mathcal{U}}$ acting on the physical matter, as follows:
\begin{equation}
    \ket{\psi(\Tilde{\mathcal{G}})}=\prod_{\site{y}\in\text{odd}} \Tilde{\theta}_{h\left(\site{y}\right)}\left(\site{y}\right)\prod_{\site{x}\in\text{even}} \theta^{\dagger}_{h\left(\site{x}\right)}\left(\site{x}\right)\ket{\psi(\mathcal{G})}\equiv \hat{\mathcal{U}}\ket{\psi(\mathcal{G})}.
\end{equation}
Hence, it is clear that $p_0(\mathcal{G})$ remains invariant under pure gauge transformations,
\begin{equation} \label{eq:matter_transformation_wave_func}
\braket{\psi\left(\Tilde{\mathcal{G}}\right)}=\bra{\psi\left(\mathcal{G}\right)}  \hat{\mathcal{U}}^\dagger\,\hat{\mathcal{U}}\ket{\psi\left(\mathcal{G}\right)}=\braket{\psi\left(\mathcal{G}\right)}.
\end{equation}

\bibliography{references.bib}

\end{document}